\documentclass[aps,preprint,showpacs ,amssymb, amsmath,pre]{revtex4}
\usepackage{bm}

\begin{document}

\title{On the properties of the Volkov solutions of the Klein-Gordon equation}

\author{Madalina Boca}
\email{madalina.boca@g.unibuc.ro}
\affiliation{Centre for Advanced Quantum Physics,  University of Bucharest, PO Box MG-11, 077125, Magurele,  Romania}

\begin{abstract}
We present an elementary proof based on a direct calculation of the property of completeness  at constant time of the solutions of the Klein-Gordon equation for a charged particle in a plane wave electromagnetic field. We also review different forms of the orthogonality and completeness relations previously presented in the literature and  we discuss the possibility to construct the Feynman propagator for the particle in a plane-wave laser pulse as an expansion in terms of Volkov solutions. We show that this leads to a rigorous justification for the expression of the transition amplitude, currently  used  in the literature, for a class of laser assisted or laser induced processes.

\end{abstract}
\pacs{03.65.Ca, 03.65.Pm}
\maketitle
\section{Introduction}\label{s-intro}
The subject of this paper is the study of some basic properties of the solutions of the Klein-Gordon equation \cite{KG1,KG2} for a charged particle in a plane wave electromagnetic field. These solutions, as well as those of Dirac equation, derived in 1935 by Volkov \cite{Volkov}, are a very useful tool in the study of the interaction of particles with intense laser fields. Also solutions for more complicated systems, as charged particles in a combination of a plane wave and a static magnetic \cite{e1} or crossed electric and magnetic fields \cite{e2} fields were studied. A very detailed review of the solutions of Dirac/Klein-Gordon equation for several forms of the external field can be found in the book  of Bagrov and Gitman \cite{BG}. The terminology used  varies in the literature: in some papers, the term ``Volkov'' is used to designate the solutions of the Dirac equation for a charged particle in a plane wave electromagnetic field and ``Gordon-Volkov'' for the  case of the Klein-Gordon equation; more often, the term Gordon-Volkov is used to designate the non-relativistic limit of the Volkov states. Sometimes only the name Volkov is used for both  cases: spin one half and spinless particle  \cite{nev};  we shall adopt  this simplified terminology,  mentioning explicitly the particle spin when necessary.

The study of Volkov  solutions is motivated by their use in the description of laser induced or laser assisted scattering processes in the high intensity regime. In such a description the charged particle dressed by the laser is described by a Volkov function, which allows the non-pertubative treatment of the field. Early descriptions of processes as non-linear Compton scattering, laser assisted Mott scattering, laser assisted Compton scattering, and laser assisted bremsstrahlung use this formalism for the case of a monochromatic laser field. For a review of the literature on these topics see \cite{rev}. A merit of the Volkov solutions is that they can be written also for the case of plane-wave pulse with finite duration; this model was used in recent studies of non-linear Compton scattering \cite{BF-Compton,Heinzl}, or more exotic processes as pair creation in light by light scattering \cite{BH}. The approach common to all these studies is based  on the use of a perturbation theory with Volkov states as the unperturbed basis set.  Another topic involving the use of Volkov solutions is a relativistic generalization \cite{KM1,KM2} of the Krammers-Henneberger transformation for the case of laser atom interaction, based on the use of  an operator constructed from Volkov and free solutions of the Dirac equation. In all this type of calculations, the orthogonality and completeness properties of the Volkov states are essential for the development of the theory and they were assumed as true.

In this paper, using a method similar to that employed in a previous   calculation for the the Dirac case \cite{BF-Dirac}, we present proofs based on direct calculation of the relations of orthogonality and completeness at constant time for the Volkov solutions of the Klein-Gordon equation. We also show that using these properties one can give a derivation, based on the ${\cal S}$ matrix formalism,  of the expression of the transition amplitude  for the case of a charged spinless particle in the presence of an intense plane wave electromagnetic field and of another potential. We think an explicit  proof is useful as in  previous publications the transition amplitude  expression was written directly, being assumed tacitly as valid.

In Sec. \ref{s-volkov} we introduce the notations used in this paper, we write the Klein-Gordon Volkov states and different forms of the orthogonality and completeness relations previously used in the literature;  the derivation of the solutions is briefly presented in Appendix \ref{A-C-V}. In Sec. \ref{s-ortho}  we  present a proof of the orthogonality with respect to the Klein-Gordon inner product at constant time, and  in Sec. \ref{s-compl} we derive two relations which  express  the completeness property at constant time;  we use three identities previously proven and employed in a similar calculation referring to the Volkov solutions of the Dirac equation \cite{BF-Dirac}. Finally, in Sec. \ref{s-green} we show that, using the orthogonality and completeness properties at constant time, a Feynman-type propagator can be constructed as a superposition of Volkov solutions, and a perturbation theory with Volkov functions as the unperturbed basis states can be built. 

\section{The Volkov solutions of the Klein-Gordon equation}\label{s-volkov}

In classical electromagnetism a plane-wave laser field propagating along the direction of unit vector ${\bf n}$ is described by a vector potential ${\bf A}$ and a scalar potential $\Phi$, both of them depending  on time and coordinates only through the combination $\phi\equiv ct-{\bf n}\cdot {\bf r}$. The corresponding four-vector of the electromagnetic field is 
\begin{equation}\label{A}
A(\phi)\equiv(\frac{\Phi(\phi)}{c};{\bf A}(\phi)),
\end{equation} 
and the Lorentz gauge condition implies that  $A(\phi)\cdot n=0$, where $n$ is the notation for the four vector $n\equiv(1,{\bf n})$; with this, the argument $\phi$ of $A$ is written as a four product 
\begin{equation}
\phi=n\cdot x,\quad  x\equiv(ct, {\bf r}).\label{def-phi}
\end{equation}
The results presented in this paper are valid for both the monochromatic case, and  for the more general case of a finite pulse, obeying the condition
\begin{equation}
\lim\limits_{\phi\rightarrow\pm\infty}\Phi(\phi)=0,\quad \lim\limits_{\phi\rightarrow\pm\infty}{\bf A}(\phi)={\bf 0}.\label{f-p}
\end{equation}
For the following calculations it is convenient to introduce also another four-vector $\widetilde n\equiv(1,-{\bf n})$ and the variable 
\begin{equation}
\widetilde \phi=\widetilde n\cdot x\equiv ct+{\bf n}\cdot{\bf r};\label{def-tilde-phi}
\end{equation}
other notations we shall use are ${\bf b}_{\|}$ and ${\bf b}_{\perp}$ for the components parallel and respectively orthogonal to the laser propagation direction of an arbitrary vector ${\bf b}$: ${\bf b}_{\|}={\bf n}({\bf n}\cdot{\bf b})$, ${\bf b}_{\perp}={\bf n}\times({\bf b}\times{\bf n})$. 

The Klein-Gordon Hamiltonian for a particle of mass $m$ and electric charge $e<0$ in the laser field $A(\phi)$ is, 
\begin{equation}
H(x)=-c^2\Pi_{\mu}\Pi^{\mu}+m^2c^4,\quad \Pi_{\mu}={\mathrm i}\hbar\partial_{\mu}-eA_\mu(\phi).\label{KG-ham}
\end{equation}
and the Volkov solutions of the corresponding equation 
\begin{equation}
H(x)\psi(x)=0,\label{KG-eq}
\end{equation}
are
\begin{equation}
\psi_{\pm}(p;x)=\frac{1}{\sqrt{2E}(2\pi\hbar)^{3/2}}\exp\left\{\mp\frac{\mathrm i}{\hbar}p\cdot x\pm\frac{\mathrm i}{2\hbar(n\cdot p)}\int\limits^{\phi}_{\phi_0}d\chi\left[e^2A^2(\chi)\mp2eA(\chi)\cdot p\right]\right\}.\label{VS-pm-fin}
\end{equation}
They depend on the momentum $p\equiv(\sqrt{(mc)^2+{\bf p}^2},{\bf p})$, with ${\bf p}\in{\bf R}^3$; the constant $\phi_0$ is arbitrary, as explained in Appendix \ref{A-C-V}.  The subscript $\pm$  is used to indicate the connection with the free particle case: the Volkov solutions with the subscript $+/-$ reduce to the free Klein-Gordon plane-wave of momentum $p$ and positive/negative energy in two cases: i) in the zero field limit, and ii) in the case of an arbitrary finite pulse (\ref{f-p}) in the limit $t\rightarrow-\infty$, for any finite ${\bf r}$. In the description of the processes in the presence of a laser pulse with finite duration, the second case  gives the possibility to impose simple initial conditions. Although sometimes, for simplicity the solutions (\ref{VS-pm-fin}) are named ``positive/negative energy Volkov states'', the meaning is that they {\it originate}, each, from a positive/negative energy free solution.

We mention here that the equivalent problem in the case of the Dirac equation has similar solutions: one obtains for each value  $p$ four bispinors $\psi_i(p;x)$ $i=1,\ldots,4$, two of them corresponding to positive energy free solutions, and the other two to negative energies. The spinor part depends on the four momentum and on the vector potential of the electromagnetic field, and it is multiplied by the same phase as in the spinless case.

One can easily check that $\psi_{\pm}(p;x)$  are eigenvectors of the operators ${\bf P}_{\perp}=-{\mathrm i}\hbar {\bm{\nabla}}_{\perp}$ and $n\cdot P={\mathrm i}\hbar (\frac{\partial}{\partial (ct)}+\frac{\partial}{\partial ({\bf n}\cdot{\bf r})})$, with the eigenvalues $\pm{\bf p}_{\perp}$ and $\pm(n\cdot p)$ respectively.

The  Klein-Gordon scalar product at constant time  for a particle in the electromagnetic field described by the vector potential (\ref{A}) is \cite{Schweber}
\begin{eqnarray}
\langle f(x),g(x)\rangle&=&\int d{\bf r}f^*(x)\left[{\mathrm i}\hbar\overset{\leftrightarrow}{\partial_0} -2ceA_0(\phi)\right]g(x)\nonumber\\
&\equiv&\int d{\bf r}\left[f^*(x){\mathrm i}\hbar\frac{\partial g(x)}{\partial t}-g(x){\mathrm i}\hbar \frac{\partial f^*(x)}{\partial t}-2ceA_0(\phi)f^*(x)g(x)\right]\,.\label{sp-KG}
\end{eqnarray}
With this definition, the orthogonality relations  at constant time for the Volkov solutions are
\begin{equation}
\langle\psi_{\pm}(p_1;x),\psi_{\pm}(p_2;x)\rangle=\pm\delta({\bf p}_1-{\bf p}_2),\quad \langle\psi_{+}(p_1;x),\psi_{-}(p_2;x)\rangle=0;\label{O-KG}
\end{equation}
we shall present the proof in the next section.
The general form of the completeness relation for a set of functions $\{\phi_i\}$ orthogonal with respect to  an inner product, 
\begin{equation}
\langle \phi_i,\phi_k\rangle=d_i\delta_{ik},
\end{equation}
is
\begin{equation}
{\bf I}=\sum\limits_i \frac1{d_i}\phi_i\langle\phi_i,\cdot\rangle
\end{equation}
where ${\bf I}$ is the unit operator. If we use the previous equation for the case of Klein-Gordon Volkov states, one must replace the general sum over the index $i$ by an integral over the parameter ${\bf p}$ and a sum over the two types of solutions, of positive and  respectively negative energies, and the coefficient $d_i$ by $+1$ and, respectively, $-1$. With these, we have
\begin{equation}
 {\bf I}=\int d{\bf p}\psi_+(p)\langle\psi_+(p),\cdot\rangle-\int d{\bf p}\psi_-(p)\langle\psi_-(p),\cdot\rangle
\end{equation}
and, using the definition  (\ref{sp-KG}) of the Klein-Gordon scalar product one sees that the above relation is equivalent to the pair of relations 
\begin{equation}
{\cal I}_1\equiv\int d{\bf p}\left[\psi_{+}(p;x)\psi_{+}^*(p;x')-\psi_{-}(p;x)\psi_{-}^*(p;x')\right]=0\label{C1}
\end{equation}
and
\begin{eqnarray}
{\cal I}_2&\equiv&\int d{\bf p}\left[\psi_{+}(p;x)\left({\mathrm i}\hbar\frac{\partial \psi_{+}(p;x')}{\partial t}\right)^*-\psi_{-}(p;x)\left({\mathrm i}\hbar\frac{\partial \psi_{-}(p;x')}{\partial t}\right)^*\right.\label{C2}\\
&-&\left.2ceA_0(\phi')\left(\psi_+(p;x)\psi_+^*(p;x')-\psi_-(p,x)\psi_-^*(p;x')\right)\vphantom{\left({\mathrm i}\hbar\frac{\partial \psi_{+}(p;x')}{\partial t}\right)^*}\right]=\delta({\bf r}-{\bf r}')\,;\nonumber  
\end{eqnarray}
we emphasize that in the above relations the position four vectors are taken for the same time $x=(ct,{\bf r})$, $x'=(ct,{\bf r}')$. Note that if the relation (\ref{C1}) is obeyed, then the expression of ${\cal I}_2$ reduces to the simpler form
\begin{equation}
{\cal I}_2=\int d{\bf p}\left[\psi_{+}(p;x)\left({\mathrm i}\hbar\frac{\partial \psi_{+}(p;x')}{\partial t}\right)^*-\psi_{-}(p;x)\left({\mathrm i}\hbar\frac{\partial \psi_{-}(p;x')}{\partial t}\right)^*\right].\label{C2-s}
\end{equation}

 In the remaining of this section we shall briefly mention several previous  papers presenting  different forms of the orthogonality and completeness properties of the Volkov solutions. The importance of these properties for the construction of a perturbation theory using the Volkov functions as the unperturbed basis set was early recognized by Neville and Rohrlich \cite{nev}. They proposed the ``null plane'' formulation of the electrodynamics; for the case of Klein Gordon equation and a laser pulse characterized by a vector potential ${\bf A}(\phi)\perp{\bf n}$ they introduced the ``null plane'' coordinates,  ${\bf r}_{\perp},\,u,\,v$, with $u$  and $v$ the same as the variables $\phi$ and  $\widetilde\phi$   we have  defined in Eqs. (\ref{def-phi}) and (\ref{def-tilde-phi}), and the scalar product on the hyperplane $u=\,{\mathrm{constant}}$ 
\begin{equation}
\langle f(x),g(x)\rangle_{u}=-{\mathrm i}\hbar\int d{\bf r}_{\perp}dvf^*(x)\overset{\leftrightarrow}{\partial}_{v}g(x).
\end{equation} In the mentioned papers it is also discussed the possibility to construct a complete orthogonal set of wave-packets of Volkov solutions.  

Gitman {\it et al} \cite{G-np} gave the generalization of the null plane scalar product for several cases of electromagnetic field consisting in a superposition of a transverse plane-wave and longitudinal electric or magnetic field. They also derived for these fields the corresponding generalized Volkov solutions of the Dirac and Klein-Gordon equation which originate from positive energy states, and proved by direct calculation the orthogonality property
\begin{equation}
\langle\psi_{\pm}(p_1,x),\psi_{\pm}(p_2,x)\rangle_{u}=\delta({\bf p}_1-{\bf p}_2).
\end{equation}

 The proof of the orthogonality with respect to the scalar product at constant time in the spin one half case 
\begin{equation}
\int d{\bf r}\psi_i^{\dag}(p_1;{\bf r},t)\psi_j(p_2;{\bf r},t)=\delta_{ij}\delta({\bf p}_1-{\bf p}_2)
\end{equation}
 was given by Ritus \cite{Ritus}, and subsequently rederived by Filipowicz \cite{Fil} and  Zakowickz \cite{Zack}. In the case of Klein-Gordon equation the similar property is stated, without a proof, in the book Bagrov and Gitman \cite{BG}.

The completeness property was less discussed; for the Klein-Gordon case  Gitman {\it et al} \cite{G-np} presented a completeness relation which, for the case of a transverse plane-wave and Klein Gordon equation reduces to 
\begin{equation}
{\mathrm i}\hbar \frac{\partial}{\partial v}\int d{\bf p}\psi_{+}(p,{\bf r}_{\perp},v,u)\psi_{+}(p,{\bf r}',v',u)=\delta({\bf r}_{\perp}-{\bf r}'_{\perp})\delta(v-v'),
\end{equation}
they also gave an analogous relation for the Dirac case. Also for the case of spin 1/2 completeness relations valid on the null plane were given by Ritus \cite{Ritus-Ann} and Bergou and Varro \cite {BV}. The completeness at constant time for the Dirac case
\begin{equation}
\sum_i\int d{\bf p}\psi_i(p;{\bf r},t)\psi^{\dag}(p;{\bf r}',t)=\delta({\bf r}-{\bf r}')\label{CD}
\end{equation}
 is stated without a proof in the book by Bagrov and Gitman \cite{BG}; it was recently derived   by Boca and Florescu \cite{BF-Dirac}.

In the following two sections we present a proof of the properties of orthogonality, Eq. (\ref{O-KG}) and completeness,  Eqs.(\ref{C1}) and  (\ref{C2-s}), at constant time for the Klein-Gordon case; part of the calculation is similar to that presented previously in \cite{BF-Dirac} for the case of spin 1/2.

\section{The orthogonality of the Volkov solutions}\label{s-ortho}

We discuss  separately two cases.

 i) {\it The  two Volkov solutions originate from   plane waves with the same sign of energy} and momenta $\,p_1$ and respectively $\,p_2$.

 We use the explicit expression (\ref{VS-pm-fin}),  in the scalar product (\ref{sp-KG})  in which  the integral over ${\bf r}_{\perp}$ can be directly calculated leading to a two dimensional delta function $\delta({\bf p}_{1\perp}-{\bf p}_{2\perp})$.  This way we are left with only an one-dimensional integral over the variable  ${\bf n}\cdot{\bf r}$
\begin{eqnarray}
\langle\psi_{\pm}(p_1;x),\psi_{\pm}(p_2;x)\rangle&=&\pm c\frac{1}{(4\pi\hbar)}\frac1{\sqrt{4E_1E_2}}\delta({\bf p}_{1\perp}-{\bf p}_{2\perp})\left(n\cdot p_1+n\cdot p_2\right)\times\label{ort-1}\\
&&\hspace*{-2cm}\int\limits_{-\infty}^{\infty}d({\bf n}\cdot {\bf r})\left[1+\frac{({\bf p}_{1\perp}\mp e{\bf A}_{\perp}(\phi))^2+(mc)^2}{(n\cdot p_1)(n\cdot p_2)}\right]\times\nonumber\\
&&\hspace*{-2cm}\exp\left\{\mp\frac{\mathrm i}{2\hbar}\left(n\cdot p_1-n\cdot p_2\right)\left[\widetilde \phi-\frac1{(n\cdot p_1)(n\cdot p_2)}\int\limits^{\phi}_{\phi_0}d\chi\left[({\bf p}_{1\perp}\mp e{\bf A}_{\perp})^2+(mc)^2\right]\right]\right\}.\nonumber
\end{eqnarray}
The previous expression can be easily calculated if one makes a change of variable
\begin{equation}
{\bf n}\cdot{\bf r}\longrightarrow \zeta\equiv\widetilde \phi-\frac1{(n\cdot p_1)(n\cdot p_2)}\int\limits^{\phi}_{\phi_0}d\chi\left[({\bf p}_{1\perp}\mp e{\bf A}_{\perp})^2+(mc)^2\right],
\end{equation}
whose Jacobian is exactly the factor in front of the exponential in the equation (\ref{ort-1})
\begin{equation}
\frac{d\zeta}{d({\bf r}\cdot{\bf n})}=1+\frac{({\bf p}_{1\perp}\mp e{\bf A}_{\perp}(\phi))^2+(mc)^2}{(n\cdot p_1)(n\cdot p_2)}.
\end{equation}
Thus we obtain 
\begin{eqnarray}
\langle\psi_{\pm}(p_1;x),\psi_{\pm}(p_2;x)\rangle&=&\frac{\pm c}{(4\pi\hbar)}\delta({\bf p}_{1\perp}-{\bf p}_{2\perp})\frac{n\cdot p_1+n\cdot p_2}{\sqrt{4E_1E_2}}\int \limits_{\phi_0}^{\infty}d\zeta{\mathrm e}^{\frac{\mathrm i}{2\hbar}(n\cdot p_1-n\cdot p_2)\zeta}\nonumber\\
&=&\pm\frac{n\cdot p_1}{E_1}\delta({\bf p}_{1\perp}-{\bf p}_{2\perp})\delta(n\cdot p_1-n\cdot p_2)=\pm\delta({\bf p}_1-{\bf p}_2)
\end{eqnarray}
which completes the proof of orthogonality and normalization for the case i).

ii)  {\it The  two Volkov solutions originate from   plane waves with different  signs of the energy} and  momenta $\,p_1$ and respectively $\,p_2$.

The scalar product between a positive energy Volkov solution and a negative energy one is treated  similarly   to the case i). We have
\begin{eqnarray}
\langle\psi_{-}(p_1;x),\psi_{+}(p_2;x)\rangle&=& c\frac{1}{(4\pi\hbar)}\frac1{\sqrt{4E_1E_2}}\delta({\bf p}_{1\perp}+{\bf p}_{2\perp})\left(n\cdot p_1-n\cdot p_2\right)\times\label{ort-2}\\
&&\hspace*{-2cm}\int\limits_{-\infty}^{\infty}d({\bf n}\cdot {\bf r})\left[1-\frac{({\bf p}_{1\perp}+e{\bf A}_{\perp}(\phi))^2+(mc)^2}{(n\cdot p_1)(n\cdot p_2)}\right]\times\nonumber\\
&&\hspace*{-2cm}\exp\left\{-\frac{\mathrm i}{2\hbar}\left(n\cdot p_1+n\cdot p_2\right)\left[\widetilde \phi+\frac1{(n\cdot p_1)(n\cdot p_2)}\int\limits^{\phi}_{\phi_0}d\chi\left[({\bf p}_{1\perp}+ e{\bf A}_{\perp})^2+(mc)^2\right]\right]\right\};\nonumber
\end{eqnarray}
as in the previous case, the change of variable
\begin{equation}
{\bf n}\cdot{\bf r}\longrightarrow \zeta\equiv\widetilde \phi+\frac1{(n\cdot p_1)(n\cdot p_2)}\int\limits^{\phi}_{\phi_0}d\chi\left[({\bf p}_{1\perp}+ e{\bf A}_{\perp})^2+(mc)^2\right],
\end{equation}
leads to
\begin{equation}
\langle\psi_{+}(p_1;x),\psi_{-}(p_2;x)\rangle=\frac{n\cdot p_1}{E_1}\delta({\bf p}_{1\perp}+{\bf p}_{2\perp})\delta(n\cdot p_1+n\cdot p_2)=0.
\end{equation}
In the above equation the last equality holds because the four product $n\cdot p$ is positive for any four-momentum $p$.
\section{The completeness of the Volkov solutions}\label{s-compl}

We present in the following a proof based on direct calculation of the two relations (\ref{C1}) and (\ref{C2-s}). In both of them, we first separate the integral in two parts, one corresponding to the positive and  the other to  the negative energy solutions, and, in the integral over the negative energy solutions, we make the change of variable ${\bf p}_{\perp}\rightarrow-{\bf p}_{\perp}$.  We use the notations
\begin{equation}
a=\frac{\phi-\phi'}{2\hbar}=\frac{\widetilde \phi'-\widetilde\phi}{2\hbar}, \quad b=\frac1{2\hbar}\int\limits_{\phi'}^{\phi}d\chi[({\bf p}_{\perp}-e{\bf A}_{\perp}(\chi))^2+(mc)^2];\label{not-c}
\end{equation}
we mention that, as the integrand in the definition (\ref{not-c}) of $b$ is positive, we have $ab>0$ for any $\phi$ and $\phi'$. With these notations, the integral in Eq. (\ref{C1}) becomes
\begin{equation}
{\cal I}_1=\frac1{2(2\pi\hbar)^3}{\mathrm e}^{-\frac{\mathrm i}{\hbar}\int\limits_{\phi'}^{\phi}eA_0(\chi)d\chi}\int d{\bf p}_{\perp}{\mathrm e}^{\frac{\mathrm i}{\hbar}{\bf p}_{\perp}\cdot({\bf r}-{\bf r}')}\int\limits_{-\infty}^{\infty}\frac{d({\bf n}\cdot{\bf p})}{E}\left[{\mathrm e}^{-i[(n\cdot p)a-\frac{b}{(n\cdot p)}]}-{\mathrm e}^{i[(n\cdot p)a-\frac{b}{(n\cdot p)}]}\right].
\end{equation}
Next we follow the same steps as in the case of the  proof of the completeness relation for the Dirac equation \cite{BF-Dirac}: in the integral over $({\bf n}\cdot {\bf p})$  we make the change of variable
\begin{equation}
{\bf n}\cdot{\bf p}\rightarrow v=(n\cdot p),\quad dv=c\frac{(n\cdot p)}Ed({\bf n}\cdot{\bf p})\label{var-v}
\end{equation}
with the result
\begin{equation}
{\cal I}_1=\frac{\mathrm i}{c}\frac1{(2\pi\hbar)^3}{\mathrm e}^{-\frac{\mathrm i}{\hbar}\int\limits_{\phi'}^{\phi}eA_0(\chi)d\chi}\int d{\bf p}_{\perp}{\mathrm e}^{\frac{\mathrm i}{\hbar}{\bf p}_{\perp}\cdot({\bf r}-{\bf r}')}\int\limits_0^{\infty}\frac{dv}v\sin(av-\frac{b}v)
\end{equation}
The integral over $v$ in the previous relation was already discussed in \cite{BF-Dirac}, and it was proven that is vanishes for $ab>0$ (Appendix A of the cited paper):
\begin{equation}
\int\limits_0^{\infty}\frac{dv}v\sin(av-\frac{b}v)=\lim\limits_{\epsilon\rightarrow 0, L\rightarrow\infty}S(a,b,\epsilon,L)=0
\end{equation}
where 
\begin{equation}
S(a,b,\epsilon,L)=\int\limits_{\epsilon}^{L}\frac{dv}v\sin(av-\frac{b}v)\,;
\end{equation}
then we have
\begin{equation}
{\cal I}_1=0\label{res-C1}
\end{equation}
which ends the first part of the proof.

In order to prove the identity (\ref{C2-s}) we use  the  same  notations (\ref{not-c}) and make the same changes of variables as before  (${\bf p}_{\perp}\rightarrow -{\bf p}_{\perp}$ in the integral over the negative energy solutions, and the one defined Eq. (\ref{var-v})) and  we obtain
\begin{eqnarray}
{\cal I}_2&=&\frac1{2(2\pi\hbar)^3}{\mathrm e}^{-\frac{\mathrm i}{\hbar}\int\limits_{\phi'}^{\phi}eA_0(\chi)d\chi}\int d{\bf p}_{\perp}{\mathrm e}^{\frac{\mathrm i}{\hbar}{\bf p}_{\perp}\cdot({\bf r}-{\bf r}')}\times\\
&\times&\left\{\int\limits_0^{\infty}dv\cos(av-\frac bv)\left[1+\frac{({\bf p}_{\perp}-e{\bf A}_{\perp}(\phi'))^2+(mc)^2}{v^2}\right]+2{\mathrm i}eA(\phi')\int\limits_0^{\infty}\frac{dv}v\sin(av-\frac bv)\right\}\nonumber
\end{eqnarray}
In the  previous equation, the second integral over $v$  is equal to zero, as shown before; in the remaining integral we recognize the two functions 
\begin{equation}
C_n(a,b,\epsilon,L)=\int\limits_{\epsilon}^{L}dvv^n\cos(av-\frac bv),\quad n=0,-2
\end{equation}
 met in Eq. (60) of \cite{BF-Dirac}
\begin{equation}
\int\limits_0^{\infty}dv\cos(av-\frac bv)=\lim\limits_{\epsilon\rightarrow 0,L\rightarrow\infty}C_0(a,b,\epsilon,L)=\pi\delta(a),
\end{equation}
\begin{equation}
\int\limits_0^{\infty}dv\frac{\cos(av-\frac bv)}{v^2}=\lim\limits_{\epsilon\rightarrow 0,L\rightarrow\infty}C_{-2}(a,b,\epsilon,L)=\pi\delta(b)\,.
\end{equation}
Using them, we have 
\begin{equation}
{\cal I}_2=\frac{\pi}{2(2\pi\hbar)^3}\int d{\bf p}_{\perp}{\mathrm e}^{\frac{\mathrm i}{\hbar}{\bf p}_{\perp}\cdot({\bf r}-{\bf r}')}\left\{\delta(a)+\delta(b)[({\bf p}_{\perp}-e{\bf A}_{\perp}(\phi'))^2+(mc)^2]\right\}=\delta({\bf r}-{\bf r}')\,.\label{res-C2}
\end{equation}
 The last equality  was obtained  with the identities
\begin{equation}
\delta(a)\equiv\delta(\frac{\phi-\phi'}{2\hbar})=2\hbar\delta(z-z')\label{i1}
\end{equation}
and
\begin{equation}
\delta(b)\equiv\delta(\frac1{2\hbar}\int\limits_{\phi'}^{\phi}d\chi[({\bf p}_{\perp}-e{\bf A}(\chi))^2+(mc)^2])=\frac{2\hbar}{[({\bf p}-e{\bf A}(\phi))^2+(mc)^2]}\delta(z-z')\label{i2}
\end{equation}
The results (\ref{res-C2}) and (\ref{res-C1}) end the proof of the completeness at constant time of the Volkov solutions of the Klein-Gordon equation.

\section{The transition amplitude for a particle in an intense laser pulse under the influence of an external perturbation}\label{s-green}

 In this section we justify the expression of the transition amplitude between Volkov  states for a spinless charged particle subject to the action of an intense  electromagnetic plane wave field  and to another weak perturbation described by the interaction Hamiltonian $H_{\mathrm{int}}(x)$.  

  If the intensity of the laser field is very large, such that it can not be treated perturbatively, the  approach  used previously in the literature, for example for the non-linear Compton scattering \cite{Ehl,Panek} on spinless particles,  was to   write directly the transition amplitude as  the matrix element of the perturbation between Volkov states. For instance, in  the case of two Volkov solutions with positive energy, one writes 
\begin{equation}
{\cal A}_{if}=-\frac{\mathrm i}{c\hbar}\int d^4x\psi_+^*(p_2;x)H_{\mathrm{int}}(x)\psi_+(p_1;x)\,.\label{A-if}
\end{equation}
This expression,  formally identical to that valid for the case of a free particle under the action of the perturbation $H_{\mathrm{int}}(x)$, can be understood intuitively, particularly for   the  monochromatic plane wave which extends indefinitely. As the Volkov states reduce to free plane-waves in the absence of the external field, the expression is used also in the pulse case, when the particles are free at the beginning and at the end of the pulse. In this section we present a rigorous  justification  of (\ref{A-if}), based on the ${\cal S}$ matrix theory.

In order to prove  (\ref{A-if}) we start from the Feynman propagator for the case of a charged spinless particle in a plane wave electromagnetic field,   built from Volkov states
\begin{equation}
K(x,x')=-{\mathrm i}\hbar\theta(t-t')\int d{\bf p}\psi_+(p,x)\psi_+^*(p;x')-{\mathrm i}\hbar\theta(t'-t)\int d{\bf p}\psi_-(p;x)\psi_-^*(p;x')\,;\label{KG-F}
\end{equation}
  The action of $K(x,x')$  on an arbitrary solution $\phi(x)$ of the Klein-Gordon equation for the particle in a plane-wave electromagnetic pulse is obtained in the same formal way as for a free particle \cite{Greiner}. Given  the completeness property of the Volkov solutions,  $\phi(x)$ can be written as a superposition of them 
\begin{equation}
\phi(x)=\int d{\bf p}\,c_+({\bf p})\psi_{+}(p;x)+\int d{\bf p}\,c_-({\bf p})\psi_{-}(p;x)\equiv \phi_+(x)+\phi_-(x)
\end{equation}
where the two components $\phi_{\pm}(x)$ are the parts of the wave-function $\phi(x)$ {\it originating} from positive, respectively negative energy plane waves (see the discussion in Sec \ref{s-volkov}). The action of $K(x',x)$ on $\phi(x')$ is written using the  Klein-Gordon scalar product (\ref{sp-KG})
\begin{equation}
K(x',x)\phi(x)=\int d{\bf r}K(x',x)[{\mathrm i}\hbar\overset{\leftrightarrow}{\partial}_0-2ceA_0(\chi)]\phi(x),
\end{equation} 
and, with the expression (\ref{KG-F}) of the propagator and the properties of orthogonality and completeness  one obtains 
\begin{equation}
K(x',x)\phi(x)=-{\mathrm i}\hbar\theta(t'-t)\phi_+(x')+{\mathrm i}\hbar\theta(t-t')\phi_-(x')
\end{equation}
which means that $K(x',x)$ propagates forward in time any function $\phi_+(x')$ {\it originating} from positive energy solutions, and backward in time the functions $\phi_-(x')$ {\it originating} from negative energy solutions.  The role played by  the  Feynman propagator of the Klein-Gordon equation for the charged particle in a plane-wave electromagnetic field is  identical to that of the free propagator in the free particle case.

Using the properties of the Volkov solutions, one can easily see that $K(x',x)$  is also a Green function of the Klein-Gordon equation.  Indeed, using the Hamiltonian (\ref{KG-ham}) and the expression (\ref{KG-F}) of $K(x,x')$, one obtains
\begin{eqnarray}
&&H(x)K(x,x')=\nonumber\\
&&=-{\mathrm i}\hbar\int d{\bf p}\left[\theta(t-t')H(x)\psi_+(p,x)\psi_+^*(p;x')+\theta(t-t')H(x)\psi_-(p;x)\psi_-^*(p;x')\right]+\nonumber\\
&&+\left[-{\mathrm i}\hbar^3\delta'(t-t')+2ce\hbar^2A_0(\phi)\delta(t-t')\right]\int d{\bf p}\left[\psi_+(p,x)\psi_+^*(p;x')-\psi_-(p;x)\psi_-^*(p;x')\right]-\nonumber\\
&&-\hbar^2\delta(t-t')\int d{\bf p}\left[{\mathrm i}\hbar\frac{\partial\psi_+(p,x)}{\partial t}\psi_+^*(p;x')-{\mathrm i}\hbar\frac{\partial\psi_-(p;x)}{\partial t}\psi_-^*(p;x')\right]
\end{eqnarray}
The first integral in the above equation vanishes, as $\psi_{\pm}(p;x)$ are solutions of the Klein-Gordon equation, the second row is also equal to zero since it is proportional to the integral ${\cal I}_1$, defined in Eq. (\ref{C1}); in the third integral we recognize the complex conjugate of the integral ${\cal I}_2$ in its simplified form (\ref{C2-s}), and using the result (\ref{res-C2}) we obtain
\begin{equation}
H(x)K(x,x')=-\hbar^2\delta(t-t')\delta({\bf r}-{\bf r}')
\end{equation}
which ends the proof. 

Now, we  come back to the case of the charged particle interacting with a plane-wave laser field and subject  to   another interaction  described by $H_{\mathrm{int}}(x)$. In the Hamiltonian 
\begin{equation}
H=-c^2\Pi_{\mu}\Pi^{\mu}+m^2c^4+H_{\mathrm{int}}(x)\equiv H_0+H_{\mathrm{int}}(x)\label{h-tot}
\end{equation}
we consider  as the unperturbed system  the particle in the laser field and  $H_{\mathrm{int}}(x)$ is treated as a perturbation; this approach is appropriate for the case of very intense laser fields.
 We follow exactly the same steps as in the case of the free particle under the action of $H_{\mathrm{int}}$, assumed of finite duration,
\begin{equation}
\lim_{t\rightarrow\pm\infty}H_{\mathrm{int}}(x)=0
\end{equation}
 but use Volkov states instead of free plane-waves; the ${\cal S}$ matrix element between the initial state and a Volkov state of positive energy and momentum  $p_2$ is
\begin{equation}
{{\cal S}_{if}=\langle\psi_+(p_2;x),\psi_i(x)\rangle\vphantom{\frac12}\vline}_{t\rightarrow\infty}.
\end{equation}
In the previous equation $\psi_i(x)$ is the solution of the Klein-Gordon equation with the Hamiltonian (\ref{h-tot}) which evolves from the initial state, assumed a Volkov state of positive energy and momentum $p_1$ state at $t\rightarrow-\infty$; in fact, as mentioned in Sect. \ref{s-volkov}, for the case of a finite laser pulse, $\psi_i(x)$ reduces  at $t\rightarrow-\infty$ to a plane wave free solution. The solution $\psi_i(x)$  obeys the integral equation
\begin{equation}
\psi_i(x)=\psi_+(p_1;x)+\frac1{\hbar^2c}\int d^4yK(x,y)H_{\mathrm{int}}(y)\psi_i(y)
\end{equation}
with $K(x,y)$ is the Green function (\ref{KG-F}) of the Klein-Gordon equation,  analyzed  before. In  the first order perturbation theory with respect to $H_{\mathrm{int}}(x)$, $\psi_i(x)$ is approximated as
\begin{equation}
\psi_i(x)\approx\psi_+(p_1;x)+\frac1{\hbar^2c}\int d^4yK(x,y)H_{\mathrm{int}}(y)\psi_+(p_1;y)
\end{equation}
and the ${\cal S}$ matrix becomes
\begin{equation}
{\cal S}_{if}\approx\delta({\bf p}_1-{\bf p}_2)+{\cal A}_{if}
\end{equation}
with 
\begin{equation}
{\cal A}_{if}=\frac1{\hbar^2c}\int d^4x'\psi_+(p_1;x')H_{\mathrm{int}}(x')\lim\limits_{t\rightarrow\infty}\langle\psi_+(p_1;x),K(x,x')\rangle.
\end{equation}
Using in the previous equation the expression (\ref{KG-F}) of the propagator and the properties of the Volkov solutions one obtains the expression (\ref{A-if}) which ends the rigorous justification of the formula of the transition amplitude for a charged particle dressed by an intense laser field and under the influence of the perturbation $H_{\mathrm{int}}$.

\appendix
\section{Derivation of the Volkov solutions of the Klein-Gordon equation}\label{A-C-V}
 Here we present briefly the derivation of the Volkov solutions for the spinless particle; the calculation is similar to that presented by Brown and Kibble \cite{BK} or Neville and Rohlich \cite{nev}. We make in the Klein-Gordon equation (\ref{KG-eq}) the change of variables
\begin{equation}
\{t,{\bf r}\}\rightarrow\{\phi,\widetilde\phi,{\bf r}_{\perp}\} 
\end{equation}
 with $\phi=ct-{\bf r}\cdot {\bf n}$ and $\,\widetilde \phi=ct+{\bf r}\cdot {\bf n}$ already defined in Sect. II. This   leads to the equation
\begin{equation}
4\hbar^2\frac{\partial^2\psi(\phi,\widetilde\phi,{\bf r}_{\perp})}{\partial\phi\partial\widetilde\phi}+2{\mathrm i}\hbar e (A(\phi)\cdot \widetilde n)\frac{\partial\psi(\phi,\widetilde\phi,{\bf r}_{\perp})}{\partial \widetilde\phi}+\left[({\bf P}_{\perp}-e{\bf A}_{\perp}(\phi))^2+(mc)^2\right]\psi(\phi,\widetilde\phi,{\bf r}_{\perp})=0.\label{eKG-nv}
\end{equation}
We look for  a particular solution of the above equation of the form
\begin{equation}
\psi(\phi,\widetilde\phi,{\bf r}_{\perp})=N\exp(\frac{\mathrm i}{\hbar}{\bf r}_{\perp}\cdot{\bf p}_{\perp})\exp(-\frac{\mathrm i}{\hbar}\lambda\widetilde\phi)F(\phi);
\end{equation}
where $\lambda$ is an arbitrary {\it positive} constant and ${\bf p}_{\perp}$ is an arbitrary real vector orthogonal to ${\bf n}$. The equation obeyed by $F(\phi)$
\begin{equation}
4{\mathrm i}\hbar\lambda\frac{dF}{d\phi}=\left[({\bf p}_{\perp}-e{\bf A}_{\perp}(\phi))^2+2\lambda eA(\phi)\cdot \widetilde n+(mc)^2\right]F(\phi)
\end{equation}
has the solution
\begin{equation}
F(\phi)=\exp\left[-\frac{\mathrm i}{4\hbar\lambda}\int\limits^{\phi}_{\phi_0}d\chi[({\bf p}_{\perp}-e{\bf A}_{\perp}(\chi))^2+2\lambda eA(\phi)\cdot \widetilde n+(mc)^2]\right];
\end{equation}
in the above expression the lower limit of the integral $\phi_0$ is arbitrary, as a change of it leads only to a irrelevant phase factor.  Finally, the solution $\psi(x)$ is
\begin{equation}
\psi(x)=N\exp\left\{\frac{\mathrm i}{\hbar}\left[{\bf p}_{\perp}\cdot{\bf r}_{\perp}-\lambda\widetilde \phi-\frac{1}{4\lambda}\int\limits^{\phi}_{\phi_0}d\chi[({\bf p}_{\perp}-e{\bf A}_{\perp}(\chi))^2+2\lambda eA(\phi)\cdot \widetilde n+(mc)^2]\right]\right\};
\end{equation}
In order to show the relation between the Volkov solutions and the free Klein-Gordon plane waves we introduce a four-vector $p=(\frac{E}c,{\bf p})$ of four-norm $(mc)$,  such that the orthogonal component of ${\bf p}$ is just the vector ${\bf p}_{\perp}$ used before, and  $\lambda=\frac12(n\cdot p)>0$. Expressed in terms of the new parameter $p$, the Volkov solution become, up to a phase factor depending on $\phi_0$
\begin{equation}
\psi_+(p;x)=N\exp\left\{-\frac{\mathrm i}{\hbar}p\cdot x+\frac{\mathrm i}{2\hbar(n\cdot p)}\int\limits^{\phi}_{\phi_0}d\chi\left[e^2A^2(\chi)-2eA(\chi)\cdot p\right]\right\},\label{VS-pos}
\end{equation}
which in the zero field limit reduces to the free Klein-Gordon plane-wave of momentum $p$ and positive energy; the connection with a positive energy free solution is indicated by the subscript $+$ introduced in the previous equation. In the case of a finite laser pulse, obeying Eq. (\ref{f-p}), the simplest choice for the lower limit of the integral in (\ref{VS-pos}) is $\phi_0\rightarrow-\infty$.

The Volkov functions which reduces in the zero field limit to the negative energy Klein-Gordon plane-wave are obtained by looking for solutions of the equation (\ref{eKG-nv}) of the form 
\begin{equation}
\psi(\phi,\widetilde\phi,{\bf r}_{\perp})=N\exp(-\frac{\mathrm i}{\hbar}{\bf r}_{\perp}\cdot{\bf p}_{\perp})\exp(\frac{\mathrm i}{\hbar}\lambda\widetilde\phi)F(\phi)
\end{equation}
the final result being
\begin{equation}
\psi_-(p;x)=N\exp\left\{\frac{\mathrm i}{\hbar}p\cdot x-\frac{\mathrm i}{2\hbar(n\cdot p)}\int\limits^{\phi}_{\phi_0}d\chi\left[e^2A^2(\chi)+2eA(\chi)\cdot p\right]\right\}\label{VS-neg}.
\end{equation}
 For both signs of the energy, we choose the normalization constant, as in the free case as 
\begin{equation}
N=\frac1{\sqrt{2E}(2\pi\hbar)^{3/2}}.
\end{equation} 

\acknowledgments
This work was supported by the strategic grant POSDRU/89/1.5/S/58852, Project ``Postdoctoral programme for training scientific researchers'' cofinanced by the European Social Found within the Sectorial Operational Program Human Resources Development 2007-2013. The author is grateful to V. Florescu for the critical reading of the manuscript and useful discussions.


\begin{thebibliography}{100}
\bibitem{KG1} Gordon W 1926 {\it Z. Physik} {\bf 40} 127
\bibitem{KG2} Klein O 1927 {\it Z. Physik} {\bf 41} 407
\bibitem{Volkov} Volkov D M 1935 {\it Z. Physik} {\bf 94} 250
\bibitem{e1} Redmond P J 1965 {\it J. Math. Phys.} {\bf 6} 1163
\bibitem{e2} Lam L 1971 {\it J. Math. Phys.} {\bf 12} 299
\bibitem{BG} Bagrov V G and  Gitman D M 1990 {\it Exact solutions of relativistic wave equation} (Kluwer)
\bibitem{nev}Neville F A and Rohrlich J 1971 {\it Phys. Rev. D} {\bf 3} 1692
\bibitem{rev} Ehlotzky F, Krajewska K, Kaminski J Z 2009 {\it Rep. Prog. Phys.} {\bf 72} 046401
\bibitem{BF-Compton} Boca M and Florescu V 2009 {\it Phys. Rev. A} {\bf 80} 053403
\bibitem{Heinzl} Heinzl T, Seipt D and Kampfer D 2010 {\it Phys. Rev. A} {\bf 81} 022125
\bibitem{BH} Di Piazza A, Lotstedt E, Milstein A I and Keitel C H 2009 {\it Phys. Rev. Lett.} {\bf 103} 170403
\bibitem{KM1} Krstic P S and Mittleman M H 1990 {\it Phys. Rev. A} {\bf 42} 4037\bibitem{KM2}Krstic P S and Mittleman M H 1992 {\it Phys. Rev. A} {\bf 45} 6514
\bibitem{BF-Dirac} Boca M and  Florescu V 2010 {\it Rom. J. Phys.} {\bf 55} 511
\bibitem{Schweber} Schweber S S 1961 {\it An introduction to the relativistic quantum field theory}, (Evanston IL; Row Peterson)
\bibitem{G-np}Gitman {\it et al} 1976 {\it Rus. J. Phys.} {\bf 18} 1097 
\bibitem{Ritus} Ritus V I 1979 {\it Trudy FIAN} {\bf 111} 5 (in Russian)
\bibitem{Fil} Filipowicz P 1985 {\it J. Phys. A} {\bf 18} 1675
\bibitem{Zack} Zakowicz S 2005 {\it J. Math. Phys.} {\bf 46} 032304
\bibitem{Ritus-Ann}Ritus V I 1972 {\it Ann. Phys.} {\bf 69} 555
\bibitem{BV} Bergou J and  Varro S 1980 {\it J. Phys. A} {\bf 13} 2823
\bibitem{Panek} Panek P {\it et al} 2002 {\it Phys. Rev. A} {\bf 65} 022712
\bibitem{Ehl} Ehlotzky F 1989 {\it J. Phys. B: At. Mol. Phys.} {\bf 22} 1989
\bibitem{Greiner} Greiner R 2003 {\it Quantum electrodynamics} (Springer)
\bibitem{BK}Brown L S and Kibble T W B 1964 {\it Phys. Rev.} {\bf 133} A705 
\end{thebibliography}
\end{document}